\documentclass[prd,twocolumn,superscriptaddress,floatfix,amsmath,amssymb,amsfonts,nofootinbib,longbibliography]{revtex4-2}

\usepackage{float} 
\usepackage{scalerel}
\usepackage[normalem]{ulem}
\usepackage[english]{babel}
\usepackage{graphicx}% Include figure files
\usepackage{dcolumn}% Align table columns on decimal point
\usepackage{bm}% bold math
\usepackage{blindtext}
\usepackage{verbatim}
\usepackage{relsize}
\usepackage{mathrsfs}
\usepackage{musicography}
\usepackage{amsmath}
\usepackage{blindtext}
\usepackage{cancel}
\usepackage{physics}
\usepackage{epstopdf}
\usepackage{mathtools}
\usepackage{blindtext}
\usepackage{tensor}
\usepackage{color}
\usepackage[usenames,dvipsnames]{pstricks}
\usepackage{epsfig}
\usepackage{pst-grad} % For gradients
\usepackage{pst-plot} % For axes
\usepackage{hyperref}
\usepackage{verbatim}
\usepackage{slashed}
\usepackage{dsfont}

\newcommand{\mf}{\mathsf}%\mathsf is too long

\newcommand{\ii}{\mathrm{i}}

\renewcommand{\r}{\hat{\rho}}

\newcommand{\tc}[1]{\textsc{#1}}

\newcommand{\trr}[1]{\textcolor{red}{#1}}
\newcommand{\tbb}[1]{\textcolor{black}{#1}}

\allowdisplaybreaks[1] %<<<<<<<<<<<<<<<<<<<<<<
\begin{document}

\title{Particle Detectors from Localized Quantum Field Theories}

\author{T. Rick Perche}
\email{trickperche@perimeterinstitute.ca}

\affiliation{Perimeter Institute for Theoretical Physics, Waterloo, Ontario, N2L 2Y5, Canada}
\affiliation{Department of Applied Mathematics, University of Waterloo, Waterloo, Ontario, N2L 3G1, Canada}
\affiliation{Institute for Quantum Computing, University of Waterloo, Waterloo, Ontario, N2L 3G1, Canada}

\author{Jos\'e Polo-G\'omez}
\email{jpologomez@uwaterloo.ca}

\affiliation{Perimeter Institute for Theoretical Physics, Waterloo, Ontario, N2L 2Y5, Canada}
\affiliation{Department of Applied Mathematics, University of Waterloo, Waterloo, Ontario, N2L 3G1, Canada}
\affiliation{Institute for Quantum Computing, University of Waterloo, Waterloo, Ontario, N2L 3G1, Canada}

\author{Bruno de S. L. Torres}
\email{bdesouzaleaotorres@perimeterinstitute.ca}\affiliation{Perimeter Institute for Theoretical Physics, Waterloo, Ontario, N2L 2Y5, Canada}
\affiliation{Institute for Quantum Computing, University of Waterloo, Waterloo, Ontario, N2L 3G1, Canada}
\affiliation{Department of Physics and Astronomy, University of Waterloo, Waterloo, Ontario, N2L 3G1, Canada}

\author{Eduardo Mart\'in-Mart\'inez}
\email{emartinmartinez@uwaterloo.ca}

\affiliation{Perimeter Institute for Theoretical Physics, Waterloo, Ontario, N2L 2Y5, Canada}
\affiliation{Department of Applied Mathematics, University of Waterloo, Waterloo, Ontario, N2L 3G1, Canada}
\affiliation{Institute for Quantum Computing, University of Waterloo, Waterloo, Ontario, N2L 3G1, Canada}

\begin{abstract}
    We present a fully relativistic model for localized probes in quantum field theory. Furthermore, we show that it is possible to obtain particle detector models from localized quantum field theories that interact with a free quantum field. In particular, a particle detector model is obtained when one traces out over inaccessible degrees of freedom of the localized field. This gives rise to a particle detector model, that is, a quantum degree of freedom that couples to a free field theory in an extended region of spacetime. Moreover, we show that the predictions of traditional particle detector models and fully relativistic localized fields completely coincide to leading order in perturbation theory. %We also discuss the origin of the non-localities of non-relativistic particle detector models from a quantum field theoretical perspective.  %that the coupling of fields in an extended region of spacetime is the physical origin of the causality violations in particle detectors. %This relationship between detectors and localized quantum field theories also holds in the case where more than one detector is coupled to the field: particle detectors yield exactly the same leading order predictions as modes of two localized quantum fields coupled to a free field. This result proves that entanglement harvesting is a phenomenon in quantum field theory, and is not a consequence of non-locality or non-relativistic effects in effective models.
\end{abstract}

\maketitle

\section{Introduction}

    Quantum field theory (QFT) is one of the most successful frameworks of theoretical physics. Its predictions are among the most accurate experimental results of physics~\cite{EMfluctuating,Gyros}, and have led to insights about the fundamental building blocks of the Universe. Moreover, quantum field theory can be made compatible with general relativity in numerous regimes, allowing us to study extreme systems, such as black holes, and the physics of the early universe~\cite{HawkingRadiation,Hawking1975,birrell_davies,kayWald,mukhanov_winitzki_2007,Wald2,fulling_1989}. 
    
    Although the applications of QFT are vast, there are still many cases of interest where we do not know yet how to successfully apply this framework to address fundamental questions. For instance, as of today there is no description for a hydrogen atom entirely within QFT. This is because handling bound states in quantum field theory requires non-perturbative techniques, which are limited. Moreover, in the context of Relativistic Quantum Information, it is necessary to describe operations that are localized in space and time. \tbb{This has sparked significant interest in how to model localized operations---and more specifically, measurements which extract local information from a quantum field---in a way that is consistent with other foundational principles of relativistic QFTs such as causality and covariance~\cite{hellwig1969,Hellwig1970formal,Hellwig1970operations,Sorkin,impossibleRevisited,IanJubb}.} 

   One way of implementing physically motivated local operations within the framework of QFT is to consider that the field is measured by probes described by localized quantum systems which are not internally relativistic. The fact that these localized probes are non-relativistic allows for a simple description of localized measurements on the field. It also provides a framework to model the flow of classical and quantum information from field degrees of freedom to experimentally observable probe systems. These non-relativistic probes are usually described by so-called particle detector models (see e.g.~\cite{Unruh1976,DeWitt}). Particle detectors have found many applications in studies of quantum field theory both in flat and curved spacetimes. For instance, they have been used to study the Unruh effect~\cite{Unruh-Wald,JormaHowThermal,waitUnruh,edunruh} and Hawking radiation~\cite{Unruh1976,JormaHawking,BenitoJormaUnruhState,bhDetectorsBTZ}, as well as directly measuring the field's correlation functions~\cite{pipo,geometry}, and providing a physically motivated measurement scheme in quantum field theories~\cite{jose,hectorjose}. Particle detectors have also been used to implement numerous quantum information protocols, such as quantum collect calling~\cite{Jonsson2,collectCalling,PRLHyugens2015}, quantum energy teleportation~\cite{teleportation,Hotta2011,teleportation2014,nichoTeleport}, entanglement harvesting~\cite{Valentini1991,Reznik1,reznik2,Pozas-Kerstjens:2015,Salton:2014jaa,Pozas2016,HarvestingSuperposed,Henderson2019,bandlimitedHarv2020,ampEntBH2020,ericksonNew,mutualInfoBH,SchwarzchildHarvestingWellDone,threeHarvesting2022,twist2022,cisco2023harvesting} and noise-assisted information transmission~\cite{Katja,kojiTeleportBetter}\tbb{, and they have been widely successful in reproducing setups of experimental significance, e.g., in quantum optics~\cite{eduardoOld,eduardo,richard} and superconducting circuits~\cite{superconducting,SwitchQED,SwitchQEDUpTheLadder}}.

% One of the most widely used particle detector models is the Unruh-DeWitt (UDW) detector model~\cite{Unruh1976,DeWitt}. 

   % Entanglement harvesting is a protocol in which particle detectors couple to the field aiming to extract entanglement from it. The entanglement in the detectors can then be used to quantify the entanglement in the field between the regions where the detectors couple to. In order to ensure that the detectors are extracting the entanglement from the field, rather than communicating through the field, it is usual to consider detectors that interact in spacelike separated regions. This simple approach to quantifying entanglement in a quantum field theory has the advantage of being simple to implement in both flat and curved spacetimes~\cite{mutualInfoBH,more}, as well as being related to realistic physical systems that can couple to a quantum field~\cite{Pozas2016,HarvestingDelocalized,carol,Boris}.

    Although particle detectors have been proven to be useful tools to study fundamental properties in quantum field theories, they also have their drawbacks. Specifically, the non-relativistic nature of the detectors can make them either non-local or singular in many scenarios. After all, particle detectors are effective models, and the effects of the non-localities they display impose clear regimes of validity to their application. Even though the limits of these regimes of validity are well understood~\cite{us2,pipoFTL,generalPD,mariaPipoNew}, it has been argued that non-relativistic particle detectors may not provide an entirely satisfactory effective model for a more fundamental fully relativistic quantum-field theoretical description of realistic localized measurements~\cite{FewsterVerch,fewster2,impossibleImpossible,IanJubb,singularOrNonlocal,BeiLokCharis,BeiLokCharisBis,fewster3}.
    
    %the non-localities introduced by particle detector models can lead one to question up to what extent their predictions are consequences of the non-relativistic description, or actually correspond to physical processes. %beg the question of to what extent  whether these particle detectors are good descriptions of experimentally realizable measurements on the field while still being compatible with the relativistic framework.  %compatibility with relativity that the 
    %predictions derived using particle detectors. %has been used to challenge some of the results derived using particle detectors~\cite{max,patricia}. One of the goals of this paper is to address these criticisms, and to show that entanglement harvesting is also a prediction of a fully relativistic theory.

    %It has been argued that particle detectors cannot be a satisfactory effective model for a subjacent fully relativistic QFT description of a real measurment.

    %For instance, the non-relativistic nature of the detectors make them non-local, which can lead to controllable violations of causality. Being effective models, the causality violations arising from particle detectors is not fundamental, but rather impose clear regimes of validity to the theory. Even though these regimes are well understood [39–42], the non-localities introduced by UDW-like models has been used to challenge some of the results derived using particle detectors, in particular, spacelike entanglement harvesting [43, 44] .
    
   \tbb{An alternative to the non-local models of particle detectors is to use quantum field theory itself to describe the probes that measure the target quantum field. The measurement theory resulting from this approach, including a covariant update rule for the target field's state, was fully developed using an algebraic formalism by Fewster and Verch~\cite{FewsterVerch}, and was further studied in~\cite{fewster2,singularOrNonlocal,fewster3}. Being formulated completely within QFT, this framework is fully local and safe from any causality issues, and has been used, e.g., to provide a fundamental solution to Sorkin's impossible measurements problem~\cite{Sorkin,impossibleRevisited,impossibleImpossible}. However, it has been argued that the algebraic approach has two main pending issues~\cite{DanPragmatic,MariaDoreen2023}: first, in this formalism the problem of measuring a quantum field is addressed with a probe that is also described as a quantum field, without specifying how should the measurement of the probe field be modelled. Second, since realistic probes cannot be modelled using free field theories~\cite{maxReply}, connecting the algebraic approach with actual experiments requires in principle a treatment of localized bound states that is currently unavailable in quantum field theory. This is because, as we mentioned before, bound states in interacting field theories only arise non-perturbatively, and thus their fundamental description is likely to require sophisticated techniques that we are still missing.}

   \tbb{Here, we work with a simpler (yet still fully relativistic) description for localized quantum fields that uses an external potential to localize the degrees of freedom of the quantum field. Physically, this external potential may correspond to an external agent which localizes a system, in the same way a Coulomb field sourced by a proton would localize the electron field around it. This technique will allow us to consider a fully local quantum field theory which is also localized in space. The localized field can then be made to interact with a free field theory, thus acting as a fully relativistic localized probe.} \textcolor{black}{Notice that, of course, this construction can be subject to the same criticism as the algebraic approach since, after all, both target and probe are described as quantum fields. Indeed, the main objective of this paper is to take a further step towards the resolution of this kind of criticism of the formalism developed by Fewster and Verch~\cite{FewsterVerch} by relating the algebraic approach to particle detector models in a detailed and explicit manner.}

   \tbb{Using quantum fields localized by external potentials as fully relativistic particle detectors is the main goal of this manuscript. In exploring this idea, we will show that each mode of the localized field theory has exactly the same behaviour of a particle detector model to leading order in the coupling strength, hence allowing us to derive particle detector models from a more fundamental theory. The use of modes of localized quantum fields as particle detectors dates back to~\cite{Unruh1976}, and the idea of singling out individual modes of quantum fields to model detectors using the algebraic formalism of~\cite{FewsterVerch} was explored in~\cite{max} for the case of a free field. \textcolor{black}{Although the relation between the algebraic formalism and the detector-based approach has been mentioned in~\cite{FewsterVerch}, and the formalism of~\cite{FewsterVerch,fewster2,singularOrNonlocal,fewster3} naturally allows for the introduction of an external potential, a clear connection between the two approaches to model local operations in QFT has not yet been explicitly explored in a concrete scenario}. In this work, we provide a setup that explicitly allows to connect both frameworks, thus bridging the rigour of the algebraic formulation of local operations in QFT with the success of the particle detector approach at modelling the physics relevant to experiments in a laboratory setting. Moreover, having a derivation of particle detector models from a first-principles Relativistic QFT framework may help these models be studied within the algebraic perspective of quantum field theory in the future. This may provide new insights into the quantum information protocols in QFT which so far have only been studied using detector-based approaches.}

   %We also analyze the case where \emph{two} localized quantum fields are used to probe a free quantum field theory. We also find that in this context, any two modes of each of the localized QFTs behave exactly like two particle detectors, to leading order in the coupling strength. In particular, by considering explicit examples, we show that two localized quantum fields can extract entanglement from a free field, even if their interaction regions are spacelike separated. This proves that the phenomenon of entanglement harvesting is not a consequence of non-localities present in particle detector models, and can be be implemented in fully relativistic theories, contrary to the claims of ~\cite{max,patricia}. 
    
    This manuscript is organized as follows. In Section~\ref{sec:detectors}, we review particle detector models and discuss the friction between relativistic considerations and non-relativistic particle detectors. In Section~\ref{sec:localizedQFT} we discuss the fundamental properties of localized quantum field theories. In Section~\ref{sec:QFTPD} we show that when a localized quantum field interacts with a free theory, each of its localized modes behaves like a particle detector to leading order in perturbation theory.  %Section~\ref{sec:fullHarvesting} is devoted to entanglement harvesting using two fully relativistic localized quantum fields to probe a free field. 
    The conclusions of our work can be found in Section~\ref{sec:conclusions}.

\section{Particle Detector models}\label{sec:detectors}

The goal of this section is to review particle detector models, \tbb{as well as} some of their applications and limitations. We introduce particle detectors \tbb{and their applications} in Subsection~\ref{sub:UDW}, and we discuss the known interplay between the model and causality and covariance in Subsection~\ref{sub:causality}.

%\tbb{SOMETHING ABOUT SECTION II B}

\subsection{The UDW Model and its generalizations}\label{sub:UDW}

A particle detector is a non-relativistic quantum system which locally couples to a quantum field. These models are inspired by an idea first considered by William G. Unruh in~\cite{Unruh1976}, and later used by Bryce DeWitt in~\cite{DeWitt}. The model and its generalizations have become known as Unruh-DeWitt (UDW) detectors. These models are extremely versatile and useful for studying properties of quantum fields in flat and curved spacetimes, as well as for studying aspects of quantum information in QFT. Examples of applications of UDW detectors are measuring the temperature of a quantum field~\cite{Unruh1976,HawkingGibbons,Letaw1981,Unruh-Wald,Takagi,Levin1993,JormaHowThermal,Korsbakken2004,DeBievre2006,matsasUnruh,edunruh,Hodgkinson2014,JuarezAubry2014,Ng2014,JormaHawking,VerchUnruh,waitUnruh,Garay2016,Ahmadzadegan2018,unruhEffectNoThermal,assymtoticBenito,Biermann2020,Good2020,unruhSlow,Arrechea2021,mine,DanIreneML,bunnyCircular}), measuring the correlations in a quantum field~\cite{pipo,geometry}, implementing measurements and state preparation in QFT~\cite{jose,hectorjose}, quantifying the entanglement in quantum field theory~\cite{ericksonNew,patricia,kelly}, and implementing quantum information protocols in QFT~\cite{teleportation,Jonsson2,collectCalling,PRLHyugens2015,quantClass,KojiCapacity,kojiTeleportBetter}.

In order to make the construction of particle detector models concrete, let us consider a $3+1$ dimensional\footnote{Although our results can easily be generalized to spacetimes of different dimensions, for simplicity we focus on the concrete example of $3+1$ in this manuscript.} spacetime $\mathcal{M}$ with a real scalar quantum field $\hat{\phi}(\mf x)$. The detector is modelled as a system with an internal quantum degree of freedom and undergoes a trajectory $\mf z(\tau)$ parametrized by its proper time. Its internal dynamics is described by a $\tau$-independent self-adjoint Hamiltonian $\hat{H}_\tc{d}$ with discrete spectrum.

Consider an observable $\hat\mu$ in the detector's Hilbert space that does not commute with its free Hamiltonian $\hat H_\textsc{d}$. The detector interacts linearly with the field via the operator $\hat{\mu}$, which, in the interaction picture, is written as $\hat{\mu}(\tau)$. The interaction is assumed to be localized around the detector's trajectory $\mf z(\tau)$ in a region defined by the support of a spacetime smearing function $\Lambda(\mf x)$. We prescribe the interaction in terms of a scalar interaction Hamiltonian density (also called Hamiltonian weight), which can be written as~\cite{us}
\begin{equation}\label{eq:hIUDW}
    \hat{h}_I(\mf x) = \lambda \Lambda(\mf x) \hat{\mu}(\tau(\mf x)) \hat{\phi}(\mf x),
\end{equation}
where $\tau(\mf x)$ denotes the Fermi normal time coordinate~\cite{poisson} associated with the trajectory $\mathsf{z}(\tau)$, and $\lambda$ is the coupling strength.

It is also common to assume that the detector satisfies the Fermi rigidity condition~\cite{Schlicht,JormaRigid,us,us2,generalPD}, so that the spacetime smearing function is assumed to be localized within the region where Fermi normal coordinates $(\tau,\bm x)$ associated to the trajectory $\mf z(\tau)$ are defined~\cite{us,us2,generalPD}, and factors as $\Lambda(\mf x) = \chi(\tau) f(\bm x)$. The functions $\chi(\tau)$ and $f(\bm x)$ then define the switching on and off of the interaction and the spatial shape of the detector, respectively. Under the assumption that the operator $\hat{\mu}(\tau)$ is dimensionless, the switching function is usually also assumed to be dimensionless (so that the constant switching $\chi(\tau) = 1$ would correspond to a detector switched on for an infinitely long time), and the smearing function $f(\bm x)$ has the units of a spatial density, $[f(\bm x)] = E^3$, for a given energy scale $E$. In $3+1$ spacetime dimensions, this implies that the coupling strength is dimensionless.

From this simple model, it is possible to compute the detector's state after the interaction with the field by applying time dependent perturbation theory to the detector-field system. The time evolution operator in the interaction picture can be written as
\begin{equation}\label{eq:UItau}
    \hat{U}_I = \mathcal{T}_\tau\exp\left(- \ii \int \dd V \, \hat{h}_I(\mf x)\right),
\end{equation}
where $\dd V$ is the spacetime invariant volume element and $\mathcal{T}_\tau \exp$ denotes the time ordered exponential with respect to the Fermi normal time coordinate $\tau$~\cite{us2}. Notice that when one considers microcausal\footnote{A theory is said to be microcausal if observables defined in spacelike separated regions commute. This condition ensures that there are no violations of causality in the theory~\cite{Haag}.} interactions, it is not necessary to prescribe the time parameter used in the time ordering. However, as we will discuss in Subsection~\ref{sub:causality}, the Hamiltonian density of Eq. \eqref{eq:hIUDW} does require such a specification in general, since for spatially smeared detectors the model violates the microcausality condition, albeit in a way that is controlled by the detector's smearing~\cite{mariaPipoNew}.

Expanding the time evolution operator using a Dyson series we can write
\begin{equation}\label{eq:UIDyson}
    \hat{U}_I = \openone + \hat{U}^{(1)}_I+ \hat{U}^{(2)}_I + \mathcal{O}(\lambda^3), 
\end{equation}
where
\begin{align}\label{eq:U1U2}
    \hat{U}_I^{(1)} &= - \ii \int \dd V \, \hat{h}_I(\mf x),\nonumber\\
    \hat{U}_I^{(2)} &= - \int \dd V \dd V' \, \hat{h}_I(\mf x)\hat{h}_I(\mf x')\theta(\tau(\mf x) - \tau(\mf x')),
\end{align}
and $\theta(u)$ denotes the Heaviside theta function, which is used to implement the time ordering operation. We then assume that the detector and the field are initially uncorrelated, so that the full state of the detector-field system is of the form $\hat{\rho}_0 = \hat{\rho}_\tc{d,0} \otimes \hat{\rho}_\phi$, where $\r_{\tc{d},0}$ is the detector's initial state and $\r_\phi$ denotes the initial state of the field. We further assume that the field state is a zero-mean Gaussian state (also called \textit{quasifree}), so that 1) all the even correlation functions of the field can be written in terms of the Wightman function \mbox{$W(\mf x,\mf x') = \tr\normalsize(\hat{\phi}(\mf x) \hat{\phi}(\mf x')\hat{\rho}_{\phi}\normalsize)$}, and 2) the odd point functions all vanish.

One can then explicitly write the final state of the detector $\hat{\rho}_\tc{d}$ to leading order in perturbation theory by tracing over the field's degrees of freedom. This final state is entirely given in terms of integrals of the correlation functions $W(\mf x,\mf x')$ and the smeared monopole operators $\hat{M}(\mf x) = \Lambda(\mf x) \hat{\mu}(\tau(\mf x))$. Explicitly, the final state of the detector is given by
\begin{align}
    \hat{\rho}_\tc{d} = &\hat{\rho}_{\tc{d},0} + \lambda^2 \int \dd V \dd V' \:W(\mf x, \mf x')
        \Big(\hat{M}(\mf x')\r_{\tc{d},0}\hat{M}(\mf x)\label{eq:rhoD} \\
        &\:\:\:\:\:\:\:\:\:\:\:\:\:\:\:\:\:\:\:\:\:\:\:\:\:\:\:\:\:\:\:\:\:-\hat{M}(\mf x)\hat{M}(\mf x')  \r_{\tc{d},0}\theta(\tau-\tau')\nonumber\\
        &\:\:\:\:\:\:\:\:\:\:\:\:\:\:\:\:\:\:\:\:\:\:\:\:\:\:\:\:\:\:\:\:\:\:-\r_{\tc{d},0} \hat{M}(\mf x)\hat{M}(\mf x')\theta(\tau'-\tau) \Big)\nonumber\\
        &\:\:\:\:\:\:\:\:\:\:\:\:\:\:\:\:\:\:\:\:\:\:\:\:\:\:\:\:\:\:\:\:\:\:\:\:\:\:\:\:\:\:\:\:\:\:\:\:\:\:\:\:\:\:\:\:\:\:\:\:\:\:\:\:\:\:\:\:\:\:\:+\mathcal{O}(\lambda^4).\nonumber
\end{align}
As it can be seen from above, the detector's final state now contains information about the correlation function of the field, $W(\mf x, \mf x')$, and, as mentioned before, this dependence can be used to infer properties of the field.%, such as the temperature experienced along the detector's trajectory~\cite{mine,DanIreneML,others}, or the entanglement acquired between the detector and the field~\cite{others,kelly}, among other applications~\cite{ahmed,others}.

%Although the model we have presented here is a two-level system that couples to the field, it can be easily generalized (see e.g.~\cite{Pozas2016,neutrinos,antiparticles,carol,mine,pitelli,Boris}). In fact, in~\cite{generalPD} it was shown that a localized quantum mechanical system with a wavefunction description can be described around a trajectory in a background curved spacetime. By considering a coupling with a free quantum field, the quantum system then becomes a particle detector model. %Particle detectors are then an suitable description for the interaction of localized systems (such as atoms) with an external quantum field (such as the electromagnetic field). 
%One simple generalization of the two-level particle detector introduced here would be considering any other quantum system that couples to $\hat\phi(\mf x)$ via a localized operator valued spacetime current $\hat{M}(\mf x)$ (a generalized monopole moment in the interaction picture), according to the interaction Hamiltonian
%\begin{equation}
    %\hat{h}_I(\mf x) = \lambda \hat{M}(\mf x) \hat{\phi}(\mf x).
%\end{equation}
One particular generalization of the model which will be especially relevant here is when the detector's internal degree of freedom is a harmonic oscillator of frequency $\Omega$ and the smeared monopole operator is given by\footnote{This kind of generalizations with a complex smearing function is used to model physical systems like, for example, atomic systems coupled to the electromagnetic field~\cite{Pozas-Kerstjens:2015,richard}.} \mbox{$\hat{M}(\mf x) = \Lambda(\mf x) \hat{a}(\tau)+ \Lambda^*(\mf x) \hat{a}^\dagger(\tau)$}. In this case the scalar interaction Hamiltonian density can be written as
\begin{equation}\label{eq:hIHO}
    \hat{h}_I(\mf x) = \lambda ( \Lambda(\mf x) e^{- \ii \Omega \tau}\hat{a}+ \Lambda^*(\mf x)e^{\ii \Omega \tau}\hat{a}^\dagger) \hat{\phi}(\mf x),
\end{equation}
where $\hat{a}$ and $\hat{a}^\dagger$ are the ladder operators for the harmonic oscillator. Importantly, if the detector is initially in the ground state, the model can be consistently reduced to a finite dimensional system, and for many applications, even a two-level system (see, e.g.~\cite{EricksonZero}).% Effectively, the generalized monopole moment is picked as \mbox{$\hat{M}(\mf x) = \Lambda(\mf x) (e^{- \ii \Omega \tau}\hat{a}+e^{\ii \Omega \tau}\hat{a}^\dagger)$} in order to reproduce this case.

Even though particle detector models have a diverse range of applications and have been shown to be a good approximate description for localized quantum systems in interaction with quantum fields~\cite{eduardoOld,eduardo,richard}, these models are intrinsically an effective description. This is mainly due to the fact that, when spatially smeared, the detector's internal dynamics is still described by non-relativistic quantum mechanics and not QFT. Due to this non-relativistic description, the model may be in conflict with general relativity in many different situations, which we will discuss in detail in Subsection~\ref{sub:causality}. A way to bypass these issues would be to derive the detector model from a fully relativistic quantum field theory (as has been pointed out in, e.g.,~\cite{FewsterVerch,max,maxReply}), instead of assuming a non-relativistic description for the internal detector dynamics. Although attempts have been made in this direction~\cite{FewsterVerch,max,flaminiaAchim}, these attempts have not yet been connected to the UDW family of models when these models work as good effective approximations. This is particularly relevant since particle detector models have proven to faithfully reproduce the physics of many experimental setups~\cite{eduardoOld,eduardo,richard,neutrinos,antiparticles,superconducting,SwitchQED,SwitchQEDUpTheLadder}. In order to better understand why the UDW models work in some regimes and what their true limits are, it would be very useful to derive the effective UDW model from a fully relativistic QFT. Our primary goal in this manuscript is to introduce localized detectors which are described by fully relativistic quantum fields, and discuss under which conditions we can understand the UDW family of particle detectors models as (experimentally accessible) subsystems---in the standard quantum mechanical sense of the word---of fully relativistic systems. %show how it is possible to obtain localized detectors which admit a fully relativistic description. 

\subsection{Covariance and Causality}\label{sub:causality}

As mentioned earlier, the effective non-relativistic character of the spatially smeared particle detector model can lead to inconsistencies with general relativity. However, the scales at which these inconsistencies appear are controlled by the scale of the spatial smearing of the detector and how it compares with other scales in the problem (e.g., interaction duration, or  proper acceleration). These scales in turn set the limits of validity of particle detector models---thus delimiting the range of physical scenarios for which the models produce faithful predictions. Indeed, the limits of applicability of detector models have been extensively studied~\cite{eduardoOld,eduardo,richard,us2,pipoFTL,jose,generalPD,mariaPipoNew}. The goal of this subsection is precisely to summarize the aspects of particle detector models that may lead them to be incompatible with relativity, as well as the regimes in which these incompatibilities become irrelevant, thus stipulating the regimes of validity of the models.

%We first turn our attention to the prescription of the time evolution operator from Eq. \eqref{}, which privileges the Fermi normal coordinate time associated with the detector's trajectory $\tau(\mf x)$. This has been studied in detail in~\cite{us2}. 
All inconsistencies with relativity that may arise in the employment of particle detector models can be traced back to the fact that in smeared UDW-like models the detector couples to multiple spacelike separated points. Specifically, the detector is described using a single quantum degree of freedom defined along the trajectory $\mf z(\tau)$. The interaction Hamiltonian of Eq.~\eqref{eq:hIUDW} then couples the monopole operator $\hat{\mu}(\tau)$ to the quantum field $\hat{\phi}(\mf x)$ at multiple spacelike separated points  for each value of the Fermi time coordinate $\tau$. This is also true for the generalizations of the model with complex smearing functions (e.g., Eq.~\eqref{eq:hIHO}=. That is, the detector's only degree of freedom couples simultaneously to the field at all the events in a surface of constant $\tau$ which are within the support of the spatial smearing $f(\bm x)$, and thus evolves influenced by all of them at the same time. Mathematically, this translates into the fact that the scalar interaction Hamiltonian density of the detector, $\hat{h}_I(\mf x)$, violates the  microcausality condition, i.e., it does not commute with itself when evaluated at spacelike separated events that lie within the support of the detector's spatial smearing.

One way in which the violation of the microcausality condition leads to inconsistencies with relativity is through violations of covariance. Specifically, the prescription of the time evolution operator (see Eq.~\eqref{eq:UItau}) uses a time ordered exponential that privileges a particular time parameter (for example, the Fermi coordinate $\tau$). If the microcausality condition were fulfilled by $\hat{h}_I(\mf x)$, all possible time parameters would yield the same time evolution operator, and the predictions of the model would be unambiguous. However, for spatially smeared detectors the result does depend on the time parameter chosen to perform the time ordering. The scalar interaction Hamiltonian density in Eq.~\eqref{eq:hIUDW}, though covariant itself, might not yield a covariant time evolution operator~\cite{us2}. The magnitude of the discrepancy between the evolution operators associated with different choices of time parameter is nevertheless controlled by the characteristic scales of the detector, and goes to zero as the extension of the spatial smearing goes to zero. This means that particle detectors give effectively covariant predictions provided that their proper sizes are sufficiently small.

Another way in which the violation of the microcausality condition puts detector models in conflict with relativity is through potential violations of causality. The simultaneous backreaction of the detector on the field at spacelike separated events can lead the model to predict faster-than-light propagation of information. This issue is intimately related to the measurement problem in QFT, which was first identified by Rafael Sorkin in the context of non-selective finite-rank projective measurements~\cite{Sorkin}, and has been further extended to local infinite-rank ones~\cite{impossibleRevisited} and more general operations~\cite{IanJubb}. In all these setups, there is an instantaneous operation applied simultaneously to spacelike separated points of spacetime, and (under specific circumstances~\cite{impossibleRevisited,IanJubb}) this leads to one region of spacetime signalling to another one that is spacelike separated from it. The difference between the setups involving particle detectors and those involving raw QFT operations is that, with detectors, we can precisely quantify the magnitude of the causality issues. In fact, once again, the violations are controlled by the characteristic scales of the problem (e.g., the detector size or its coupling strength)~\cite{pipoFTL,mariaPipoNew}, therefore allowing us to identify the regimes where the issues are negligible, and the models are thus employable. 

Finally, we comment on the results of~\cite{generalPD}, which present an effective description for localized non-relativistic quantum systems in curved spacetimes, and applies such a description to particle detector models. The formalism naturally imposes a restriction on the size of the systems which can be described around a given trajectory in a background spacetime. In essence, the size of the non-relativistic system must be smaller than a combination of the inverse proper acceleration of the trajectory and the curvature radius of the rest surfaces associated with the trajectory.

 %This intuition is made precise in~\cite{generalPD}, where an effective description for localized non-relativistic quantum systems in curved spacetimes is put forward and applied to particle detectors. The formalism has a natural regime of validity that restricts the size of the system in terms of its proper acceleration and the curvature of the rest surfaces of its trajectory.}

Overall, the main idea of this subsection is that, for non-pointlike detectors, detector models can be employed in those regimes where the inconsistencies with relativity are negligible. All this considered, one may wonder why we do not limit ourselves to pointlike particle detectors, for which the Hamiltonian density satisfies the microcausality condition. After all, all the conflicts with relativity come from considering non-pointlike models. The reason to consider smeared detectors is double: \mbox{1) while} pointlike detector models can be used to obtain predictions in certain scenarios, they can also lead to UV divergences~\cite{JormaRigid,Satz_2007}; and 2) the physical systems that we try to model with particle detectors (e.g., atoms~\cite{richard}, superconducting qubits~\cite{superconducting}, etc.) are certainly not pointlike objects.

\vspace{-0.3cm}

\section{Localized Quantum Fields}\label{sec:localizedQFT}

In order for a state in a quantum field theory to be spatially localized at all times, one requires a strong interaction that produces an effective confining potential, giving rise to bound states which are localized in a given region of space. Fundamentally, the strong interactions required to produce localized states should also be described by quantum field theory. For example, the hydrogen atom is a bound state that should be a consequence of the interactions of the standard model. However, in our current state of knowledge we are rarely able to describe bound systems completely within QFT. The reason for this difficulty is that bound states are usually outside of the regime of perturbation theory, where quantum field theory is mostly understood, and therefore their description is still an open problem~\cite{WeinbergQFT}.

Although we still do not know how to describe most bound states entirely within QFT, there are effective descriptions which still allow one to treat the system of interest as a quantum field. For instance, we can achieve this  by describing the strong interaction which localizes its states as an external potential. %Perhaps the simplest example of this treatment is in the case of a confined real scalar quantum field with null Dirichlet boundary conditions in the boundary of a given region $\Omega$. This problem can be reformulated by considering a Klein-Gordon field under the influence of a potential which is zero inside $\Omega$ and infinite in its complement. Although this specific potential is unphysical, it can be used to model fields in cavities with specific boundary conditions. %The resulting field is a relativistic field, but the specific shape of the region $\Omega$ can violate certain symmetries of the spacetime.
For example, it is possible to employ this treatment for the hydrogen atom by considering the electron to be a fermionic field under the influence of a Coulomb potential sourced by a pointlike proton~\cite{boundQED}. The predictions of this class of models of confined quantum field theories have no covariance or causality issues since, by construction, the effective QFTs are microcausal. %This is in spite of the fact that the potential may break some of the underlying symmetries of spacetime. In the example of the hydrogen atom, the Coulomb potential breaks space translation invariance, while preserving time translations and rotations around the center of the potential.

\vspace{-0.3cm}

\subsection{Scalar field with a confining potential}\label{sub:MinkConfined}

In order to describe localized quantum systems as quantum fields under the influence of an external \tbb{confining} potential\footnote{Notice that adding a (non homogeneous) potential makes the equation of motion not Poincar\'e invariant. \tbb{Of course, this is just an artifact of treating the physical system that confines the field as nondynamical. Poincar\'e invariance can be automatically restored once the physical entity that sources the potential is also endowed with degrees of freedom that transform nontrivially under the Poincar\'{e} group. In the approximation that the effect of the source is modelled in terms of an external potential, this just corresponds to also applying the group transformations to the potential itself.}} \tbb{$V(\mathsf{x})$}, we consider \tbb{for simplicity} the example of a real scalar quantum field in Minkowski spacetime . \tbb{The details of the more general case of curved backgrounds are given in Appendix~\ref{app:localizedQFTCS}.}

To prescribe the potential let us assume that there is a foliation of spacetime associated to a specific set of inertial coordinates $(t,\bm x)$ so that the potential can be written as $V(\bm x)$, i.e., the potential is static in that frame. Physically, we can think of these coordinates as the ones associated with a proper reference frame of the source of the potential. We consider a classical Klein-Gordon field  that is described by the Lagrangian
%\begin{equation}
%\tbb{\mathcal{L} = - \frac{1}{2} \partial_\mu\phi \,\partial^\mu \phi- \frac{m^2}{2} \phi^2 - F(V(\bm x),\phi),}
%\end{equation}
%\tbb{where $m$ is the field's mass, $F$ is} \sr{some bounded-from-below} \tbb{function of the potential and the field amplitude, and we use the $(-,+,+,+)$ metric signature convention. Here we will focus on the simplest confining case where the dependence on the field is always quadratic, so that the Lagrangian simplifies to}
\begin{equation}\label{eq:lag}
    \mathcal{L} = - \frac{1}{2} \partial_\mu\phi \,\partial^\mu \phi- \frac{m^2}{2} \phi^2 - V(\bm x) \phi^2,
\end{equation}
where $m$ is the field's mass and we use the $(-,+,+,+)$ metric signature convention. Extremization of the associated action yields the equation of motion for the field $\phi(\mf x)$ under the influence of the potential  potential $V(\bm x)$:%\too{Here and later I have changed the parentheses, let me know your opinion, I can change them back}
\begin{equation}\label{eq:KGpot}
    \big(\partial_\mu \partial^\mu - m^2-2V(\bm x)\big) \phi(\mf x) = 0.
\end{equation}
This equation is separable in the coordinates $(t,\bm x)$:
\begin{equation}\label{eq:sepKG}
    -\partial_t^2\phi(\mf x) + \big(\nabla^2 - m^2-2V(\bm x)\big) \phi(\mf x) = 0.
\end{equation}
Under the assumption that $V(\bm x)$ is a \tbb{continuous} positive confining potential (that is $V(\bm x) \to \infty$ as $|\bm x|\to \infty$)\footnote{Notice that many physical potentials are not confining in this strict sense, but for low enough energies they are \textit{effectively} confining, and admit a discrete spectrum in the relevant range of energies, so the formalism here also applies to them in these regimes (see Thm. XIII.1 in~\cite{ReedSimon4}).}, the operator $E^2(\bm x) = -\nabla^2 + m^2 + 2V(\bm x)$ will admit a positive compact inverse defined in $L^2(\mathbb{R}^3)$ (see Thm. XIII.16 in~\cite{ReedSimon4}). Let $\bm n$ be a multi-index, so that $\Phi_{\bm n}(\bm x)$ denotes the eigenfunctions and $\omega_{\bm n}^2$ denotes the eigenfrequencies of $E^2(\bm x)$:
\begin{equation}\label{eq:eigenfuncs}
    E^2(\bm x)\Phi_{\bm n}(\bm x) = \omega_{\bm n}^2 \Phi_{\bm n}(\bm x).
\end{equation}
Under the assumption that $\phi(t,\bm x)$ are square-integrable functions in space for each constant value of $t$, we can write 
\begin{equation}
    \phi(t,\bm x) = \sum_{\bm n} v_{\bm n}(t) \Phi_{\bm n}(\bm x) + \text{H.c.}
\end{equation}
Plugging the expression above into Eq. \eqref{eq:sepKG} and using Eq. \eqref{eq:eigenfuncs}, we obtain a differential equation for $v_{\bm n}(t)$:
\begin{equation}
    \ddot{v}_{\bm n}(t) = -\omega_{\bm n}^2v_{\bm n}(t).
\end{equation}
We can then expand the real classical solution for the field as
\begin{equation}
    \phi(t,\bm x) = \sum_{\bm n} \big(\alpha_{\bm n} e^{-\ii \omega_{\bm n} t}\Phi_{\bm n}(\bm x) + \alpha_{\bm n}^* e^{\ii\omega_{\bm n} t}\Phi_{\bm n}^*(\bm x)\big),
\end{equation}
where the complex numbers $\alpha_{\bm n}$ are defined by initial conditions.

In order to match the usual conventions adopted in quantum field theory, we pick the functions $\Phi_{\bm n}(\bm x)$ such that the mode functions $u_{\bm n}(\mf x) = e^{- \ii \omega_{\bm n}t} \Phi_{\bm n}(\bm x)$ are normalized with respect to the Klein-Gordon inner product:
\begin{equation}
    (\phi_1,\phi_2)_\tc{kg} = \ii \int_{\Sigma} \dd\Sigma^\mu\,\left(\phi_1^*\partial_\mu\phi_2 - \partial_\mu \phi_1^*\,\phi_2  \right),
\end{equation}
where $\Sigma$ is any Cauchy surface, $\dd\Sigma^\mu = n^\mu \dd\Sigma$, with $n^\mu$ being the future directed unit normal to $\Sigma$ and $\dd \Sigma$ is the induced volume element in $\Sigma$. That is, we normalize the functions $\Phi_{\bm n}(\bm x)$ so that
\begin{equation}
    (u_{\bm n},u_{\bm n'})_{\tc{kg}} = \delta_{\bm n, \bm n'}, %\quad (u_{\bm n},u_{\bm n'}^*)_{\tc{kg}} = 0, 
    \quad (u_{\bm n}^*,u_{\bm n'}^*)_{\tc{kg}} = -\delta_{\bm n, \bm n'}. 
\end{equation}
Picking the surfaces of constant $t$, we find that this amounts to the following normalization of $\Phi_{\bm n}(\bm x)$ in $L^2(\mathbb{R}^3)$:
\begin{equation}
    \int \dd^3\bm x |\Phi_{\bm n}(\bm x)|^2 = \frac{1}{2\omega_{\bm n}}.
\end{equation}

The fact that the potential $V(\bm x)$ is confining implies that each of the eigenfunctions $\Phi_{\bm n}(\bm x)$ is mostly localized around the minima of $V(\bm x)$, and decay to zero as $V(\bm x)$ increases. This means that the solutions to the equation of motion, Eq.~\eqref{eq:sepKG}, are mostly localized in a worldtube that contains the minima of the potential.

\begin{comment}
Besides the sublety of properly defining what strongly localized means, I find the previous statement vague and arguably wrong if we don't further constrain what we mean or in what conditions we mean it.

\color{red}
It is important to discuss in which sense the functions $\Phi_n(\bm x)$ are localized. Consider a set of linear operators $\mathcal{A} = \{\hat{A}_1,...,\hat{A}_n\}$ which act in $L^2(\mathbb{R}^3)$. We say that a function $f\in L^2(\mathbb{R}^3)$ is strongly supported to precision $\varepsilon$ in a set $U\subset \mathbb{R}^3$ with respect to $\mathcal{A}$ if 
\begin{equation}
    \langle f_{U^c}|\hat{A}|f_{U^c}\rangle<\varepsilon||f||^2 \quad \forall \:A\in \mathcal{A},
\end{equation}
where $f_{U^c}$ denotes the restriction of the function $f$ to the complement of the region $U$. It will be common for us to consider $\mathcal{A} = \{\openone,E^2(\bm x), \hat{\bm x},\hat{\bm p}\}$, where $\hat{\bm x}$ and $\hat{\bm p}$ are the standard position and momentum operators. Under suitable conditions for the potential $V(\bm x)$, each of the functions $\Phi_{\bm n}(\bm x)$ will be strongly supported in a finite region that contains the minima of $V(\bm x)$. This means that solutions to the equation of motion $\phi(\mf x)$ are strongly supported in a worldtube that contains the minima of potential.
\color{black}
\end{comment}

The quantization of this field can then be done by promoting the coefficients $\alpha_{\bm n}$ and $\alpha_{\bm n}^*$ to creation and annihilation operators $\hat{a}_{\bm n}$ and $\hat{a}_{\bm n}^\dagger$, which act on a Hilbert space containing  a vacuum state $\ket{0}$ such that $\hat{a}_{\bm n} \ket{0} = 0$ for all $\bm n$. The creation and annihilation operators satisfy the commutation relations $[\hat{a}_{\bm n},\hat{a}_{\bm n'}^\dagger] = \delta_{\bm{n},\bm{n}'}$. The quantum field $\hat{\phi}(\mf x)$ can then be represented as
\begin{equation}\label{eq:localizedField}
    \hat{\phi}(\mf x) = \sum_{\bm n} \big( \hat{a}_{\bm n} e^{-\ii \omega_{\bm n} t}\Phi_{\bm n}(\bm x) + \hat{a}_{\bm n}^\dagger e^{\ii\omega_{\bm n} t}\Phi_{\bm n}^*(\bm x) \big).
\end{equation}

There are key properties to be noted about the quantum field in Eq.~\eqref{eq:localizedField}. Notice that the field excitations are indexed by a countable index $\bm n$. This implies that the mode excitations of the form $\hat{a}^\dagger_{\bm n}\ket{0}$ are normalized states in the quantum field theory. That is, unlike in the case of non-localized quantum fields, the single mode excitations are physical states with well-defined values of relevant observables such as energy and momentum~\cite{PDclick}. Moreover, each mode $\bm n$ is localized around the same region as the functions $\Phi_{\bm n}(\bm x)$. These are both consequences of the fact that the potential $V(\bm x)$ is confining.

\begin{comment}
The countability of the set of indices has to be argued further using the property that V is confining. Also, I think enurable implies finite, which that tensor product generally isn't. 
\end{comment}

%Notice that because the field modes are strongly localized around the minima of $V(\bm x)$, so are the quantum field excitations, and it is justified to say that the field is strongly localized in a finite region. This description can be useful to describe fields under the influence of physically meaningful potentials which can be approximated to be classical, such as electrons under the influence of the Coulomb field of the nucleus in an atom~\cite{boundQED}, or a field under the influence of a quadratic potential.

%The importance of the confining external potential $V(\bm x)$ is twofold. On the one hand, it localizes each of the field modes, which gives rise to well defined and normalized single mode excitations, such as $\hat{a}_{\bm n}^\dagger \ket{0}$, which is strongly supported %\footnote{By strongly supported, we mean that given an $\epsilon>0$, there exists a compact region of space such that the expected value of a given set of relevant observables in the theory is smaller than $\epsilon$. This implies that for all practical purposes that have a precision smaller than $\epsilon$, the state is localized within a compact region.} around the minimum of the potential $V(\bm x)$. On the other hand, 

The confining nature of $V(\bm x)$ also implies that the Hilbert space representation associated to the vacuum state $\ket{0}$ decomposes as a countable tensor product of the Hilbert space associated to each mode. This allows one to perform operations such as partial tracing over modes and independently describe the dynamics of each single mode, while still obtaining a physical state for the localized field. %We will see that this property of localized fields has many useful applications, and can aid in the task of implementing localized operations in quantum field theory.

%\tbb{It is also worth noticing that the addition of a non-homogeneous potential breaks the Poincaré symmetry of the model, making the equation of motion not Poincaré invariant. This means that the quantization of the field will in general be different depending on the choice of inertial reference frame. However, this lack of invariance should not be confused with a break of covariance. The model is---as long as it is microcausal---fully covariant, and the potential inequivalences for the different quantizations are perfectly legitimate. Moreover, since the potential is modelling the physical system that confines the field, one can always recover the broken invariance by applying the group transformations corresponding to the lost symmetries to the potential itself.}

\tbb{For a detailed computation involving one of the most common textbook examples of confining potentials---the quadratic potential---see Appendix~\ref{app:harmonicQFT}.}

\subsection{Compactly supported fields}\label{sub:compactQFT}

In order to consider fields localized in a finite region of space one has to make some extra assumptions about the confining potential. Notice that if the potential $V(\bm x)$ is a smooth function, the field modes will not be compactly supported. To restrict the field to a compact region, we consider $3+1$ Minkowski spacetime, and use a potential which is only finite in a compact convex connected set $U \subset \mathbb{R}^3$ of diameter $d$, and $V(\bm x)$ is infinite\footnote{One could formally consider a one-parameter family of confining potentials $V_\epsilon(\bm x)$ such that $V_\epsilon(\bm x)\to \infty$ as $\epsilon\to 0^+$ for $\mf x \notin U$ and $V_\epsilon(\bm x)\to V(\bm x)$ as $\epsilon\to 0^+$ for $\mf x \in U$. For $\epsilon> 0$ this fits the class of models of Subsection~\ref{sub:MinkConfined}.} outside of $U$. In this case, the effect of the infinite potential is to set the value of the functions to zero outside of the region $U$, effectively enforcing Dirichlet boundary conditions on the field. Then the operator $E^2(\bm x) = -\nabla^2 + m^2 + V(\bm x)$ is defined in a dense domain of $L^2(U)$. The operator $E^2(\bm x)$ is positive and has discrete spectrum~\cite{ReedSimon4,Simon2008}, which gives it a discrete set of eigenfunctions $\Phi_{\bm n}(\bm x)$ associated to eigenvalues $\omega_{\bm n}^2$ such that
\begin{align}
    E^2(\bm x) \Phi_{\bm n}(\bm x) = \omega_{\bm n}^2 \Phi_{\bm n}(\bm x).
\end{align}
This leads us to the same solutions found in Eq. \eqref{eq:localizedField} for the classical equations of motion, with the additional property that $\Phi_{\bm n}(\bm x)$ are \emph{compactly} supported in $U$. This way, the solutions to Eq.~\eqref{eq:KGpot} have support in the worldtube $\mathbb{R}\times U$. The quantization procedure can be carried out like in the non-compactly supported case described in Subsection~\ref{sub:MinkConfined}, yielding a quantum field of the same form as in Eq. \eqref{eq:localizedField}. In this case, however, every mode excitation $\hat{a}_{\bm n}^\dagger \ket{0}$ is compactly supported by construction.

As a particular example of compactly supported field theory that is relevant in quantum optics, we briefly study a field in a perfectly reflective cavity, which can be treated using the formalism presented in this subsection. Specifically, we consider a cubic cavity $U_d = [0,d]^3$ where the potential vanishes in the region $U_d$, and is infinite everywhere else, i.e.,
\begin{equation}
    V(\bm x) = \begin{cases}
    0, \quad &\bm x \in U_d\\
    \infty , \quad &\bm x \notin U_d.
    \end{cases}
\end{equation}
The spatial solutions in this case are the eigenfunctions $\Phi_{\bm n}(\bm x)$ of the operator $-\nabla^2+m^2$ in $L^2(U_d)$, with the boundary condition $\Phi_{\bm n}(\bm x) = 0$ at the boundary of $U_d$. The solutions are 
\begin{equation}
    \Phi_{\bm n}(\bm x) = \frac{1}{\sqrt{2\omega_{\bm n}} }f_{n_x}(x)f_{n_y}(y)f_{n_z}(z),
\end{equation}
where
\begin{equation}
    f_n(u) = \sqrt{\frac{2}{d}}\sin(\frac{\pi n x}{d}),
\end{equation}
the energy eigenvalues are given by
\begin{equation}\label{eq:gapBox}
    \omega_{\bm n} = \sqrt{m^2 + \frac{\pi^2}{d^2}(n_x^2 + n_y^2 + n_z^2)},
\end{equation}
and the field takes the shape of Eq. \eqref{eq:localizedField}. %Changing the shape of the region $U_d$, and the boundary conditions could also be used to model cavities of other shapes and made of materials with different properties.

\begin{comment}
\subsection{Scalar Atoms}

\tbb{My plan here is to use scalar electromagnetism $\psi^*\psi \phi$ in order to describe an artificial scalar atom modelled by the field $\psi(\mf x)$ when one plugs in $\phi\sim 1/r$ for the field $\phi$ in the interaction above. This should be easier to solve than a fully featured atom, but could be a good model for the interactions of atoms with an external scalar electromagnetic field.} - \trr{The only issue with this is that either we will require more general particle detector models, or we will need to describe how to transform this model into a particle detector as well.}

\section{A Quantum Field as a Particle Detector}

In this section we will be concerned with the interaction of a localized quantum field interacting with a free quantum field - under the influence of no external potential. Our goal is to see how the localized quantum field can be used to probe a Klein-Gordon field. Our result will be that there are limits where a localized quantum field can effectively be treated as a localized non-relativistic quantum system. Therefore, before describing the localized quantum field, we will briefly review the formalism of particle detector models usually present in the literature.

\end{comment}

\section{Localized Quantum fields as particle detector models}\label{sec:QFTPD}

    In this section we will consider a detector modelled by a localized relativistic quantum field $\hat{\phi}_\tc{d}(\mf x)$ which will act as the probe for a free Klein-Gordon field $\hat{\phi}(\mf x)$, \tbb{following the} spirit \tbb{of} the Fewster-Verch measurement scheme~\cite{FewsterVerch,fewster2,fewster3}. 
    
    Consider a field $\hat{\phi}_\tc{d}(\mf x)$ in a $3+1$ dimensional globally hyperbolic spacetime under the influence of an external confining potential $V(\mf x)$ \tbb{(see Appendix~\ref{app:localizedQFTCS})}. In this case, there exists a discrete set of modes $u_{\bm n}(\mf x)$  such that the field $\hat{\phi}_\tc{d}(\mf x)$ can be written as
    \begin{equation}
        \hat{\phi}_{\tc{d}}(\mf x) = \sum_{\bm n} \big(\hat{a}_{\bm n} u_{\bm n}(\mf x)+\hat{a}_{\bm n}^\dagger u^\ast_{\bm n}(\mf x)\big).
    \end{equation}
    We will assume that this field couples linearly to a free Klein-Gordon field $\hat{\phi}(\mf x)$. The system of the two interacting fields can be described by the Lagrangian
    \begin{align}\label{eq:lagQFTPD}
        \mathcal{L} = -\frac{1}{2}\partial_\mu \phi_\tc{d}\partial^\mu \phi_\tc{d} - \frac{m_\tc{d}^2}{2}\phi_\tc{d}^2 - V(\mf x) \phi_{\tc{d}}^2 \\- \frac{1}{2}\partial_\mu \phi\partial^\mu \phi - \frac{m^2}{2} \phi^2\nonumber  - \lambda \zeta(\mf x) {\phi}_\tc{d} {\phi},\nonumber
    \end{align}
    where $\lambda$ is a coupling strength and $\zeta(\mf x)$ is a localized function that controls the  spacetime region where the interaction between $\hat{\phi}_\tc{d}(\mf x)$ and $\hat{\phi}(\mf x)$ happens. For convenience we assume that $\zeta(\mf x)$ is dimensionless, which implies that the coupling strength has dimensions of energy squared.%\trr{(This is exactly the example that they have in the FV-framework, apart from the external potential $V(\bm x)$, so when we do the measurement thing, it should be relatively easy to compare this case with the F.V.)}

    It is clear that the Lagrangian of Eq. \eqref{eq:lagQFTPD} defines two non-interacting quantum field theories at $\lambda = 0$, with a localized field $\hat{\phi}_\tc{d}(\mf x)$ and a free Klein-Gordon field $\hat{\phi}(\mf x)$. For small values of $\lambda$ one can use perturbation theory with the scalar interaction Hamiltonian density
    \begin{equation}
        \hat{h}_{\text{int}}(\mf x) = \lambda \zeta(\mf x) \hat{\phi}_\tc{d}(\mf x) \hat{\phi}(\mf x).
    \end{equation}
    Assuming the support of $\zeta(\mf x)$ to be localized in time, one can compute the time evolution operator associated with the interaction,
    \begin{equation}
        \hat{U} = \mathcal{T}\exp(-\ii \int \dd V \hat{h}_{\text{int}}(\mf x) ),
    \end{equation}
    where $\dd V = \sqrt{-g}\, \dd^4 x$ is the spacetime invariant volume element, and $\mathcal{T}\exp$ denotes the time ordering exponential with respect to \emph{any} time parameter. Unlike the interaction of smeared particle detector models with a free field, we now have a fully relativistic QFT of two interacting fields which, in particular, respects microcausality. 
    
    For simplicity we will look at the case where the field $\hat{\phi}$ starts in a zero-mean Gaussian state $\hat{\rho}_\phi$ and the field $\hat{\phi}_\tc{d}$ starts in its vacuum state $\ket{0_\tc{d}}$, so that the initial state of the two-field system is \mbox{$\hat{\rho}_0 = \ket{0_\tc{d}}\!\!\bra{0_\tc{d}}\!\otimes\hat{\rho}_\phi$}. Importantly, due to the discrete energy levels of the localized quantum field $\hat{\phi}_\tc{d}$, its Fock space $\mathcal{F}_\tc{d}$ factors as
    \begin{equation}\label{eq:FockTensor}
        \mathcal{F}_\tc{d} = \bigotimes_{\bm n} \mathcal{H}_{\bm n},
    \end{equation}
    where $\mathcal{H}_{\bm n}$ is the Hilbert space associated to each mode. A consequence of this decomposition is that the vacuum state of the localized quantum field can be written as
    \begin{equation}\label{eq:vacTensor}
        \ket{0_\tc{d}} = \bigotimes_{\bm n} \ket{0_{\bm n}},
    \end{equation}
    where $\ket{0_{\bm n}}$ denotes the state of zero occupation number for the mode $\bm n$. We can then write
    \begin{equation}
        \mathcal{H}_{\bm n} = \text{span}\left(\{(\hat{a}_{\bm n}^\dagger)^m\ket{0_{\bm n}}\text{: } m = 0,1,...\}\right).
    \end{equation}
    
    We will also assume that we only have access to one mode of the localized field $\hat{\phi}_\tc{d}(\mf x)$ described in the subspace $\mathcal{H}_{{}_{\!\bm N}\!}$ for a given $\bm N$, which labels an eigenfrequency $\omega_{{{}_{\!\bm N}\!}}\!$ and its corresponding eigenmode $u_{{}_{{\!\bm N}\!}}(\mf x)$. Denote by $\mathcal{H}_{{}_{\!\bm N}\!}^c$ the complement of this Hilbert space in the decomposition of Eq. \eqref{eq:FockTensor}, so that the detector field's Fock space factors as $\mathcal{F}_\tc{d} = \mathcal{H}_{{}_{\!\bm N}\!}\otimes \mathcal{H}_{{}_{\!\bm N}\!}^c$. The final state that we have access to will then be given by
    \begin{equation}
        \hat{\rho}_{{}_{\!\bm N}\!} = \tr_{\phi,\mathcal{H}_{{}_{\!\bm N}\!}^c}\left(\hat{U} \hat{\rho}_0\hat{U}^\dagger\right).
    \end{equation}
    Physically, the restriction of the field to the space $\mathcal{H}_{{}_{\!\bm N}\!}$ can be realized experimentally if one only has access to excitations of the localized field with energy $\omega_{{}_{\!\bm N}\!}$. For instance, consider an electromagnetic cavity which contains photodetectors that can only measure `photons' that have energy $\omega_{{}_{\!\bm N}\!}$. Effectively, an experimentalist would only have access to the space $\mathcal{H}_{{}_{\!\bm N}\!}$, and the partial trace operation that we have performed is physically meaningful. 

    Under the assumption that the coupling strength $\lambda$ is sufficiently small, we can compute the final state $\hat{\rho}_{{}_{\!\bm N}\!}$ as a power expansion in $\lambda$ by using the Dyson expansion for $\hat{U}$,
    \begin{equation}
        \hat{U} = \openone + \hat{U}^{(1)} + \hat{U}^{(2)} + \mathcal{O}(\lambda^3),
    \end{equation}
    where
    \begin{align}
        \hat{U}^{(1)} &= - \ii \int \dd V \hat{h}_{\text{int}}(\mf x),\\
        \hat{U}^{(2)} &= - \int \dd V \dd V' \hat{h}_{\text{int}}(\mf x)\hat{h}_{\text{int}}(\mf x')\theta(t-t'),
    \end{align}
    and $\theta(t)$ denotes the Heaviside theta function, with $t$ being any time coordinate. The final state of the two fields is given by
    \begin{equation}
        \r_f = \r_0 + \r^{(1)} + \r^{(2)} + \mathcal{O}(\lambda^3),
    \end{equation}
    with
    \begin{align}
        \r^{(1)} &= \hat{U}^{(1)} \r_0 + \r_0 \hat{U}^{(1)\dagger},\\
        \r^{(2)} &= \hat{U}^{(2)} \r_0 + \hat{U}^{(1)} \r_0 \hat{U}^{(1)\dagger} + \r_0 \hat{U}^{(2)\dagger}.
    \end{align}
    Notice that because the interaction Hamiltonian is linear on $\hat{\phi}(\mf x)$ and the field starts in a zero-mean Gaussian state, we have $\tr(\hat{\phi}(\mf x) \hat{\rho}_\phi) = \langle\hat{\phi}(\mf x)\rangle_{\rho_\phi} = 0$, so that the term $\hat{\rho}^{(1)}$ does not contribute after the partial trace over the degrees of freedom of $\hat{\phi}(\mf x)$. Defining 
    \begin{equation}\label{eq:Qofx}
        \hat{Q}(\mf x) = \zeta(\mf x) \hat{\phi}_\tc{d}(\mf x),
    \end{equation}
    the term $\r^{(2)}$ can be written as
    \begin{align}
        \r^{(2)} = \lambda^2 \int \dd V \dd V'\Big(&\hat{Q}(\mf x) \hat{\phi}(\mf x)\hat{\rho}_0 \hat{Q}(\mf x') \hat{\phi}(\mf x')\\
        &-\hat{Q}(\mf x)\hat{Q}(\mf x')  \hat{\phi}(\mf x)\hat{\phi}(\mf x')\hat{\rho}_0 \theta(t-t') \nonumber\\
        &-\hat{\rho}_0 \hat{Q}(\mf x)\hat{Q}(\mf x')  \hat{\phi}(\mf x)\hat{\phi}(\mf x')\theta(t'-t) \Big)\nonumber.
    \end{align}
    Partial tracing over the free field $\hat{\phi}(\mf x)$, and using \mbox{$\hat{\rho}_0 = \ket{0_\tc{d}}\!\!\bra{0_\tc{d}}\!\otimes\hat{\rho}_\phi$}, we then obtain
    \begin{align}
        &\!\!\!\!\!\tr_\phi(\r^{(2)})\! =\!\lambda^2\!\! \int\! \dd V \dd V'W(\mf x, \mf x')
        \Big(\hat{Q}(\mf x')\ket{0_\tc{d}}\!\!\bra{0_\tc{d}}\hat{Q}(\mf x) \label{eq:midComputation}\\
        &\:\:\:\:\:\:\:\:\:\:\:\:\:\:\:\:\:\:\:\:\:\:\:\:\:\:\:\:\:\:\:\:\:\:\:\:\:\:\:\:\:\:\:\:\:\:-\hat{Q}(\mf x)\hat{Q}(\mf x')  \ket{0_\tc{d}}\!\!\bra{0_\tc{d}}\theta(t-t') \nonumber\\
        &\:\:\:\:\:\:\:\:\:\:\:\:\:\:\:\:\:\:\:\:\:\:\:\:\:\:\:\:\:\:\:\:\:\:\:\:\:\:\:\:\:\:\:\:\:\:-\ket{0_\tc{d}}\!\!\bra{0_\tc{d}} \hat{Q}(\mf x)\hat{Q}(\mf x')\theta(t'-t) \Big),\nonumber
    \end{align}
    where $W(\mf x, \mf x') = \langle\hat{\phi}(\mf x) \hat{\phi}(\mf x')\rangle_{\rho_\phi}$ is the Wightman function of the field $\hat{\phi}(\mf x)$ in the state $\hat{\rho}_\phi$. 

    The next step is to trace the result of Eq. \eqref{eq:midComputation} over the space $\mathcal{H}_{{}_{\!\bm N}\!}^c$, which we assumed to be inaccessible. In order to perform this computation, we define
    \begin{align}
        \hat{Q}_{\bm n}(\mf x) = \zeta(\mf x) \left(\hat{a}_{\bm n} u_{\bm n}(\mf x)+\hat{a}_{\bm n}^\dagger u_{\bm n}^*(\mf x)\right),
    \end{align}
    so that
    \begin{equation}
        \hat{Q}(\mf x) = \sum_{\bm n} \hat{Q}_{\bm n}(\mf x).
    \end{equation}
    Now we have
    \begin{equation}
        \tr_{\mathcal{H}_{{}_{\!\bm N}\!}^c}\left(\hat{Q}(\mf x) \hat{Q}(\mf x')\right) = \sum_{\bm n \bm m} \tr_{\mathcal{H}_{{}_{\!\bm N}\!}^c}\left(\hat{Q}_{\bm n}(\mf x) \hat{Q}_{\bm m}(\mf x')\right).
    \end{equation}
    From Eq.~\eqref{eq:vacTensor}, the vacuum $\ket{0_\tc{d}}$ can be decomposed in terms of the ground state of each mode:
    \begin{equation}
        \ket{0_\tc{d}}\!\!\bra{0_\tc{d}} = \bigotimes_{\bm n \bm m}  \ket{0_{\bm n}^\tc{d}}\!\!\bra{0_{\bm m}^\tc{d}} = \hat{\rho}_{{}_{\!\bm N\!\text{\scriptsize{,0}}}}\!\!\!\!\!\!\!\!\!\!\!\!\!\!\bigotimes_{\:\:\:\:\:\:\:(\bm n,\bm m)\neq (\bm N, \bm N)} \!\!\!\!\!\!\!\! \!\!\!\ket{0_{\bm n}^\tc{d}}\!\!\bra{0_{\bm m}^\tc{d}},
    \end{equation}
    where $\r_{{}_{\!\bm N\!\text{\scriptsize{,0}}}} = \ket{0_{{}_{\!\bm N}\!}}\!\!\bra{0_{{}_{\!\bm N}\!}}$. Noticing that each $\hat{Q}_{\bm n}(\mf x)$ term only contains one creation and one annihilation operator, we then find that for $(\bm n, \bm m) \neq (\bm N, \bm N)$,
    \begin{align}
        \tr_{\mathcal{H}_{{}_{\!\bm N}\!}^c}\Big(\hat{Q}_{\bm n}(\mf x)\hat{Q}_{\bm m}&(\mf x')\ket{0_\tc{d}}\!\!\bra{0_\tc{d}}\Big)\\& = \delta_{\bm n \bm m}\zeta(\mf x)  u_{\bm n}(\mf x) \zeta(\mf x') u_{\bm m}^*(\mf x')\r_{{}_{\!\bm N\!\text{\scriptsize{,0}}}},\nonumber
    \end{align}
    which is simply a multiple of the initial state of the mode $\bm N$. For $\bm n = \bm m = \bm N$, the trace over $\mathcal{H}_{{}_{\!\bm N}\!}^c$ simply gets rid of the components of the state in $\mathcal{H}_{{}_{\!\bm N}\!}^c$, without affecting the components in $\mathcal{H}_{{}_{\!\bm N}\!}$. 
    
    Using these results it is possible to trace Eq. \eqref{eq:midComputation} over the space $\mathcal{H}_{{}_{\!\bm N}\!}^c$, which yields
    \begin{align}
        \!\!\!\r_{{}_{\!\bm N}\!}\!&=\!\hat{\rho}_{{}_{\!\bm N\!\text{\scriptsize{,0}}}} + \lambda^2 \!\int\! \dd V \dd V'W(\mf x, \mf x')
        \Big(\hat{Q}_{{}_{\!\bm N}\!}(\mf x')\r_{{}_{\!\bm N\!\text{\scriptsize{,0}}}}\hat{Q}_{{}_{\!\bm N}\!}(\mf x) \label{eq:finalRhoQFTPD}\\
        &\:\:\:\:\:\:\:\:\:\:\:\:\:\:\:\:\:\:\:\:\:\:\:\:\:\:\:\:\:\:\:\:\:\:\:\:\:\:\:\:\:\:\:\:\:\:-\hat{Q}_{{}_{\!\bm N}\!}(\mf x)\hat{Q}_{{}_{\!\bm N}\!}(\mf x')  \r_{{}_{\!\bm N\!\text{\scriptsize{,0}}}}\theta(t-t') \nonumber\\
        &\:\:\:\:\:\:\:\:\:\:\:\:\:\:\:\:\:\:\:\:\:\:\:\:\:\:\:\:\:\:\:\:\:\:\:\:\:\:\:\:\:\:\:\:\:\:-\r_{{}_{\!\bm N\!\text{\scriptsize{,0}}}} \hat{Q}_{{}_{\!\bm N}\!}(\mf x)\hat{Q}_{{}_{\!\bm N}\!}(\mf x')\theta(t'-t) \Big)\nonumber\\
        &+\lambda^2 \r_{{}_{\!\bm N\!\text{\scriptsize{,0}}}}\sum_{\bm n\neq \bm N}\int \dd V \dd V' W(\mf x, \mf x') \zeta(\mf x)  u_{\bm n}(\mf x)\zeta(\mf x')u^*_{\bm n}(\mf x') \nonumber\\
        &\:\:\:\:\:\:\:\:\:\:\:\:\:\:\:\:\:\:\:\:\:\:\:\:\:\:\:\:\:\: \times \big( 1- \theta(t-t') - \theta(t'-t)\big) + \mathcal{O}(\lambda^4).\nonumber
    \end{align}
    Notice that the last term cancels, given that \mbox{$\theta(t-t') + \theta(t'-t) = 1$}. Also notice the similarity with the leading order result for particle detectors in Eq. \eqref{eq:rhoD}. 
    
    At this stage it is possible to reinterpret the final result by considering the following effective scalar Hamiltonian density
    \begin{equation}
        \hat{h}_{\text{eff}}(\mf x) =
        \lambda \hat{Q}_{{}_{\!\bm N}\!}(\mf x) \hat{\phi}(\mf x) 
        = \lambda \!\left(\Lambda^-(\mf x)\hat{a}_{{}_{\!\bm N}\!}  +\Lambda^+(\mf x) \hat{a}_{{}_{\!\bm N}\!}^\dagger \right)\hat{\phi}(\mf x),
    \end{equation}
    where we defined the (time evolved) spacetime smearing functions $\Lambda^-(\mf x)$ and $\Lambda^+(\mf x)$ as
    \begin{equation}
        \Lambda^-(\mf x) = \zeta(\mf x) u_{{}_{\!\bm N}\!}(\mf x), \quad \text{and} \quad \Lambda^+(\mf x) = (\Lambda^-(\mf x))^*.
    \end{equation}
    The operator $\hat{h}_\text{eff}(\mf x)$ acts on the Hilbert space of the field $\hat{\phi}(\mf x)$ and on the Hilbert space $\mathcal{H}_{{}_{\!\bm N}\!}$, which is effectively a harmonic oscillator. Defining
    \begin{equation}
        \hat{U}_{\text{eff}} = \mathcal{T}\exp\left(- \ii \int \dd V \hat{h}_{\text{eff}}(\mf x)\right),
    \end{equation}
    it is possible to show that the result for $\r_{{}_{\!\bm N}\!}$ in Eq. \eqref{eq:finalRhoQFTPD} can be rewritten as
    \begin{equation}
        \r_{{}_{\!\bm N}\!} = \tr_\phi\left(\hat{U}_\text{eff}(\r_{{}_{\!\bm N\!\text{\scriptsize{,0}}}}\otimes \r_\phi )\hat{U}_{\text{eff}}^\dagger\right) + \mathcal{O}(\lambda^4).
    \end{equation}
    That is, to second order in the coupling strength, it is possible to reproduce the final state of $\hat{\phi}_\tc{d}(\mf x)$ in the subspace $\mathcal{H}_{{}_{\!\bm N}\!}$ by considering an interaction of a harmonic oscillator with the quantum field $\hat{\phi}(\mf x)$. 

    If the spacetime is static, and the external potential is invariant under the flow of the timelike Killing vector field $\partial_t$, then it is possible to establish an even more direct comparison with particle detector models. In this case, the metric can be decomposed as in Eq.~\eqref{eq:staticmetric} and, using that $V(\mf x)$ is invariant under the flow of the Killing vector field, the solutions $u_{\bm n}(\mf x)$ decompose as $u_{\bm n}(\mf x) = e^{-\ii \omega_{\bm n}t}\Phi_{\bm n}(\bm x)$. We can then write the time-evolved spacetime smearing function $\Lambda^-(\mf x)$ as
    \begin{equation}
        \Lambda^-(\mf x) = \zeta(\mf x) e^{- \ii \omega_{{}_{\!\bm N}\!} t} \Phi_{{}_{\!\bm N}\!}(\bm x) =  e^{- \ii \omega_{{}_{\!\bm N}\!} t} \Lambda(\mf x),
    \end{equation}
    where we have defined the regular spacetime smearing function as
    \begin{equation}
        \Lambda(\mf x) \coloneqq \zeta(\mf x) \Phi_{{}_{\!\bm N}\!}(\bm x).
    \end{equation}
    The effective interaction Hamiltonian then reads
    \begin{equation}
        \hat{h}_{\text{eff}}(\mf x) = \lambda  (\Lambda(\mf x)e^{-\ii \omega_{{}_{\!\bm N}\!} t}\hat{a}_{{}_{\!\bm N}\!} + \Lambda^*(\mf x)e^{\ii \omega_{{}_{\!\bm N}\!} t}\hat{a}_{{}_{\!\bm N}\!}^\dagger)\hat{\phi}(\mf x).
    \end{equation}
    In order to fully recover the expression for a particle detector undergoing a given trajectory in this spacetime, let $\bm x_0$ be the spatial coordinates of the center\footnote{This center can always be assigned. If $V(\bm x)$ has a single global minimum the centre would be at the coordinates of the minimum. Otherwise this centre can be chosen to be the `centre of mass' of the mode~\cite{DixonI,DixonII,DixonIII}. %and the shape of the metric functions $\beta(\bm x)$ and $h_{ij}(\bm x)$%This center is defined by the minimum of the potential $V(\bm x)$ and the shape of the metric functions $\beta(\bm x)$ and $h_{ij}(\bm x)$. Alternatively, one can pick 
    %This center is defined as the point such that the first geodesic moments of $|\Phi_{{}_{\!\bm N\!}}(\bm x)|^2$ vanish.%, i.e.
    %\begin{equation}
        %\int_{\Sigma} \dd \Sigma \, \sigma^{i}(t,\bm x_0;t,\bm x) |\Phi_{{}_{\!\bm N\!}}(\bm x)|^2 = 0.
    %\end{equation}
    } of the function $\Phi_{{}_{\! \bm{N}}\!}(\bm x)$. Consider the trajectory $\mf z(\tau) = (\gamma \tau, \bm x_0)$, where $\tau$ is its proper time, and $\gamma$ is the redshift factor relative to $t$. We can now define the proper energy gap as 
    \begin{equation}\label{eq:omegaN}
        \Omega = \gamma \omega_{{}_{\!\bm N}\!},
    \end{equation}
    so that the effective interaction Hamiltonian can be written as
    \begin{equation}\label{eq:heff}
        \hat{h}_{\text{eff}}(\mf x) = \lambda \left(\Lambda(\mf x) e^{-\ii \Omega \tau}\hat{a}_{{}_{\!\bm N}\!} + \Lambda^*(\mf x) e^{\ii \Omega \tau}\hat{a}_{{}_{\!\bm N}\!}^\dagger\right)\hat{\phi}(\mf x),
    \end{equation}
    which is exactly the interaction Hamiltonian of a harmonic oscillator particle detector with energy gap $\Omega$ interacting with a scalar field $\hat{\phi}(\mf x)$ (see Eq.~\eqref{eq:hIHO}). Note that %here creation and annihilation operators $\hat{a}_{{}_{\!\bm N}\!}$ and $\hat{a}^\dagger_{{}_{\!\bm N}\!}$ here denote excitations of a mode of a quantum field, and  that 
    the units of the switching function and coupling strength could be matched with the harmonic oscillator UDW model by rebalancing\footnote{Here we have $[\lambda] = E^2$ and $[\Lambda(\mf x)] = E$, where $E$ is an energy scale. This is unlike the UDW model, where the coupling strength is dimensionless and the spacetime smearing function has units of a spatial density.} the units of the function $\zeta(\mf x)$. \tbb{Moreover,} if one is only interested in ``one-particle'' excitations in the mode $\bm N$, this allows us to consistently restrict the system and Hamiltonian density of Eq. \eqref{eq:heff} into a two-level system spanned by \mbox{$\{\ket{0_{{}_{\!\bm N}\!}},\ket{1_{{}_{\!\bm N}\!}}\}$}. This reduces to the leading order interaction of a two-level UDW detector with a scalar quantum field, which is the most widely used particle detector model in the literature.

    \tbb{It is worth remarking that this ``derivation'' of particle detectors is fundamentally different from the one presented in~\cite{Unruh-Wald}. There, particle detector models were recovered from non-relativistic quantum systems, while here we recovered them from fully relativistic QFTs.} It is also important to notice that the analogy between a localized QFT detector and the particle detector models that we find here is specific to leading order perturbation theory. Indeed, there will be discrepancies between the model with $\hat{h}_\text{eff}(\mf x)$ and the model with the interaction Hamiltonian density of the full quantum field theory interaction $\hat{h}_\text{int}(\mf x)$ to fourth or higher order in the coupling strength. %This implies that it is only to leading order that the analogy between particle detectors and modes of localized quantum fields holds, and higher order effects require quantum field theory for their description.

    We have seen how one can derive a particle detector model from a fully relativistic quantum field theory. Although a previous attempt of considering ``second quantized'' UDW detectors exists~\cite{flaminiaAchim}, the particle detectors in that case are obtained from localized states of a free massive scalar field with a very high mass. However, no state of a free massive scalar field can be truly localized, as the free propagation will cause the quantum state to spread regardless of the value of the field's mass. Here we consider energy eigenstates of the detector field's free Hamiltonian which are  localized in space and can also be static. This means that, in this approach, the detector remains localized for arbitrarily long times. The small price that we have to pay in order to obtain this localized theory is the addition of an external potential $V(\bm x)$, which is non-dynamical, and prescribed by the physical entity responsible for the localization of the field $\hat{\phi}_\tc{d}(\mf x)$.

    This derivation of particle detector models as a specific mode of a localized quantum field also allows us to see where the non-localities of the usual particle detector models come from. Although we started with two microcausal theories for the fields $\hat{\phi}_\tc{d}(\mf x)$ and $\hat{\phi}(\mf x)$, we ended up with a particle detector model, which does not respect microcausality, as discussed in Section~\ref{sec:detectors}. The reason for this is that the state of the field restricted to mode $\bm N$ represents a degree of freedom which is extended in space. Moreover, this degree of freedom is restricted to a region corresponding to the support of the mode. That is, any quantum operation performed only on this mode will affect all events within the localization of the function $\Phi_{{}_{\!\bm N}\!}(\bm x)$. This is the origin of the effective non-localities and apparent causality violations discussed in Subsection~\ref{sub:causality}. Physically, any faster-than-light predictions would be a consequence of the fact that ``measuring the energy\footnote{Or any other extended observable of $\hat{\phi}_\tc{d}$.} of the field in the mode $\bm N$'' is a non-local operation. While this operation cannot truly be implemented in a way that respects causality, it can be a very approximate description of a physical process when one has a field in a cavity, or a field in a localized region of space. The idea of tracing over infinitely many modes of the field allows for an interpretation of the causality issues of particle detectors, which is in analogy with the effective causality violations that arise when one traces over UV modes of a quantum field theory. It is well-known that restricting a quantum field theory to modes of energies below a certain cutoff introduces causality violations of the order of the inverse of the cutoff~\cite{polchinski1999effective, BurgessEffectiveFieldTheory}. That is, effective causality violations arise when tracing over modes of quantum field theories, and this is also the origin of the causality issues that arise in particle detector models. %We discuss these in more detail in Appendix~\ref{app:pathintegral}, where we explicitly trace over highly energetic modes of a localized field theory using a path integral approach.
    \tbb{In fact, this connection can be made even more concrete if we formulate the reduction of the probe field to a finite number of modes in terms of path integrals, as shown, for instance, in~\cite{QFTPDPathIntegrals}.}

    On the other hand, if we consider the whole localized quantum field theory as a probe, and restrict ourselves to causal operations, no causality violations would arise. %This approach also shows that we can build generalizations of particle detector models for which causality issues would not apparent incompatibilities with relativity are reduced at higher orders in perturbation theory by considering a multi-degree of freedom detector model, implementing scenarios where more than just a single mode can be experimentally accessed. 
    Moreover, we have chosen to trace over all modes except for one, but we could also have considered tracing over all modes except for a finite amount of them, reducing the impact of the effective causality violations at higher orders. Furthermore, if we want to remain at leading order in perturbation theory, we note that each of the detector field modes effectively behaves as a harmonic oscillator particle detector, as these evolve independently to leading order in $\lambda$.

\section{Conclusions}\label{sec:conclusions}

    %We have presented a fully relativistic model of particle detectors using localized quantum field theories.

    We have presented a fully relativistic model of particle detectors using localized quantum field theories. Using this model, we have shown how non-relativistic particle detectors can be understood as effective descriptions arising from fully relativistic localized quantum field theories. Specifically, we have shown that each mode of the localized field theory behaves exactly like a particle detector to leading order in the coupling strength. The fact that the analogy between particle detectors and localized QFTs holds to leading order in perturbation theory implies that any leading order statement about particle detectors can be made into a statement about localized modes of quantum fields \tbb{(as in~\cite{FullyRelativisticEH})}. 
    
    Since our derivation of effective models of non-relativistic particle detectors comes from fully relativistic theories, it allowed us to physically motivate the non-localities of the model. Specifically, we discussed that non-localities in detector models arise from measurements of extended field observables localized within the support of the field modes that give rise to the effective detector model. This is analogous to what happens when one traces over high energy modes of a quantum field theory, which is well known to introduce a certain degree of non-locality. 

    This work aims to set the basis to connect relativistic quantum information protocols with other fruitful research avenues to study local operations in quantum field theory. In particular, we believe it may provide a way for an algebraic quantum field theory perspective to quantum information protocols in QFT, where the fundamental condition of microcausality is satisfied, and the protocols can be looked at from a fully relativistic perspective. This work also paves the way for an explicit connection between the {detector-based} measurement schemes in QFT~\cite{jose}, and {the algebraic approaches}~\cite{FewsterVerch,fewster2,fewster3}, which are stated fully covariantly in terms of the algebraic formulation of quantum field theory. %Overall, our work allows one to derive particle detectors from a fundamental quantum field theory and to precisely quantify the regimes of validity of the model, and opens the door to a new perspective on quantum information protocols within QFT.
    
    %\textcolor{magenta}{Particle detectors can be seen as the state of localized relativistic quantum fields in a given mode which interact with a free field. This effective treatment works to leading order in the coupling strength, where most calculations with detectors are performed anyways. Basically, any statement about detectors is also a statement about localized quantum field theories.}

    %\textcolor{magenta}{Entanglement harvesting has nothing to deal with the specifics of the model, and is not caused by non-relativistic quantum mechanics or ``non-localities''. Also, particle detectors are way more realistic then just sticking a two-level system in spacetime and forcing it to interact with a quantum field---we already knew all of this, but hopefully now even a monkey can understand it.}

    %\textcolor{magenta}{This work allows AQFT to be more directly connected to RQI works, and might allow us to make the two measurement protocols compatible! (future work).}

\begin{acknowledgements}
    The authors thank Christopher J. Fewster for valuable discussions. TRP acknowledges support from the Natural Sciences and Engineering Research Council of Canada (NSERC) via the Vanier Canada Graduate Scholarship. JPG acknowledges the support of a fellowship from “La Caixa” Foundation (ID 100010434, with fellowship code LCF/BQ/AA20/11820043). JPG and BSLT acknowledge support from the Mike and Ophelia Lazaridis Fellowship. EMM acknowledges support through the Discovery Grant Program of the Natural Sciences and Engineering Research Council of Canada (NSERC). EMM also acknowledges support of his Ontario Early Researcher award. Research at Perimeter Institute is supported in part by the Government of Canada through the Department of Innovation, Science and Industry Canada and by the Province of Ontario through the Ministry of Colleges and Universities. %Perimeter Institute and the University of Waterloo are situated on the Haldimand Tract, land that was promised to the Haudenosaunee of the Six Nations of the Grand River, and is within the territory of the Neutral, Anishinaabe, and Haudenosaunee people. %As of today neither Perimeter Institute or the University of Waterloo do not plan to give the land back.
\end{acknowledgements}

\appendix

\section{Localized scalar field in a curved spacetime}\label{app:localizedQFTCS}

\tbb{In this appendix, we generalize to curved backgrounds }the treatment for localized quantum fields presented in Section~\ref{sec:localizedQFT}. Consider the case where the field is defined in a general globally hyperbolic spacetime $\mathcal{M}$ of dimension $D+1$. In this case, we do not assume that there is a foliation where the potential looks static. The Lagrangian for the field under the influence of this potential is still going to be given by Eq. \eqref{eq:lag}, now with a spacetime dependent potential. The field equation can then be written as
\begin{equation}
    (\nabla_\mu \nabla^\mu - m^2 - 2 V(\mf x))\phi(\mf x) = 0, 
\end{equation}
where $\nabla_\mu$ denotes the (torsion-free) metric-compatible covariant derivative. A relevant consequence of considering this more general case is that the equation of motion for the field will not necessarily admit a separation of variables in terms of space and time. Nevertheless, assuming that $V(\mf x)$ is confining\footnote{$V(\mf x)$ being confining means that there exists a timelike curve $\mf z(\tau)$ such that $V(\mf x) \to \infty$ when $\sigma(\mf z(\tau),\mf x)\to \infty$ for each $\tau$, where $\sigma(\mf x, \mf x')$ denotes Synge's world function~\cite{poisson} (the spacetime separation between $\mf x$ and $\mf x'$).}, it is still possible to obtain a discrete basis for the space of solutions $\{u_{\bm n}(\mf x),u_{\bm n}^*(\mf x)\}$ such that any classical solution can be written as
\begin{equation}
    \phi(\mf x) = \sum_{\bm n}(\alpha_{\bm n} u_{\bm n}(\mf x) + \alpha_{\bm n}^* u_{\bm n}^*(\mf x)).
\end{equation}
The assumption that $V(\mf x)$ is confining also implies that the functions $u_{\bm n}(\mf x)$ are strongly localized when restricted to any Cauchy surface. The quantization of this theory is analogous to the case of Minkowski spacetime, and gives the quantum field
\begin{equation}
    \hat\phi(\mf x) = \sum_{\bm n}(\hat{a}_{\bm n} u_{\bm n}(\mf x) + \hat{a}_{\bm n}^\dagger u_{\bm n}^*(\mf x)).
\end{equation}

In the case of a static globally hyperbolic spacetime, where there exists a Killing vector field $\xi$ and a foliation by Cauchy surfaces which is orthogonal to $\xi$, it is possible to separate the Klein-Gordon  equation with an external potential in a similar fashion to what we did in Minkowski spacetime. Let $t$ be the coordinate of the flow of $\xi$, then the line element can be written as
\begin{equation}\label{eq:staticmetric}
    \dd s^2 = - \beta(\bm x)^2 \dd t^2 + h_{ij}(\bm x) \dd x^i \dd x^j,
\end{equation} 
where $\{x^i\}$ are coordinates in the Cauchy surfaces orthogonal to $\xi$ and $\bm x = (x^1,\dots,x^D)$. Then, it can be shown that under the assumption that the potential $V(\mf x)$ is independent of $t$% ($\mathcal{L}_\xi V = 0$)
, the basis of solutions $u_{\bm n}(\mf x)$ can be decomposed as
\begin{equation}
    u_{\bm n}(\mf x) = e^{- \ii \omega_{\bm n} t} \Phi_{\bm n}(\bm x),
\end{equation}
where $\Phi_{\bm n}(\bm x)$ are solutions to
\begin{equation}
    \frac{\beta}{\sqrt{h}}\partial_i\left(\beta \sqrt{h}\, h^{ij} \partial_j \Phi_{\bm n}\right) - \left(\omega_{\bm n}^2 +  \beta^2(m^2 + V(\bm x))\right)\Phi_{\bm n}  = 0,
\end{equation}
and $h = \det(h_{ij})$. Notice that the equation above is indeed independent of $t$, so that the separation of variables can be performed.

\section{A field localized by a quadratic potential}\label{app:harmonicQFT}

\tbb{In this appendix we consider an explicit example of the formalism of Section~\ref{sec:localizedQFT},} where $V({\bm x})$ is a quadratic potential. Specifically, we consider inertial coordinates $(t,\bm x)$ in Minkowski spacetime and a real scalar field $\phi(\mf x)$ under the influence of the potential
\begin{equation}
    V(\bm x) = \frac{|\bm x|^2}{2\ell^4}.
\end{equation}
The parameter $\ell$ has dimensions of length, and controls the strength of the potential, with smaller values of $\ell$ corresponding to stronger potentials. The equations of motion for the field then become
\begin{equation}
    \left(\partial_\mu \partial^\mu - m^2 - \frac{|\bm x|^2}{\ell^4}\right) \phi(\mf x) = 0.
\end{equation}
It is possible to find a basis of solutions to the equations above. In fact, in Cartesian coordinates, we can solve the equation  by separation of variables. If we write \mbox{$\phi(\mf x) = e^{-\ii E t}\Phi(\bm x)$}, we obtain
\begin{equation}\label{eq:EHO}
     \left(E^2 + \nabla^2 - m^2 -\frac{|\bm x|^2}{\ell^4}\right)\Phi(\bm x) = 0.
\end{equation}
The solutions for the equation above are well known from the quantum harmonic oscillator. That is, imposing the solution to be zero at infinity, we find that $E^2 - m^2$ has to be of the form $2(n_x+n_y+n_z + \frac{3}{2})/\ell^2$ for \mbox{$n_x,n_y,n_z\in\{0,1,...\}$}. We write $\bm n = (n_x,n_y,n_z)$, so that a basis of solutions can be written as
\begin{equation}
    \Phi_{\bm n}(\bm x) = \frac{1}{\sqrt{2\omega_{\bm n}}}f_{n_x}(x)f_{n_y}(y)f_{n_z}(z),
\end{equation}
where
\begin{equation}
    f_n(u) = \frac{1}{\sqrt{2^n n!}}\frac{e^{-\frac{u^2}{2\ell^2}}}{\pi^\frac{1}{4} \sqrt{\ell}}H_n(u/\ell),
\end{equation}
and $H_n(u)$ denote the Hermite polynomials. The eigenfrequencies $\omega_{\bm n}$ are characterized by the positive values of $E$ in Eq. \eqref{eq:EHO}, and are given by
\begin{equation}\label{eq:gapHO}
    \omega_{\bm n} = \sqrt{m^2+ \frac{2}{\ell^2}\left(n_x + n_y + n_z +\frac{3}{2}\right)}.
\end{equation}
The quantization of the field using the basis of solutions $\Phi_{\bm n}(\bm x)$ yields a field operator of the form~\eqref{eq:localizedField}, where the index $\bm n$ is given by $\bm n = (n_x,n_y,n_z)$.

At this stage it is possible to analyze the spatial localization of the modes of the quantum field. We see that all modes are exponentially decaying as $e^{-|\bm x|^2/2\ell^2}$. That is, the parameter $\ell$ which is inversely related to the strength of the potential $V(\bm x)$ is also related to the spatial localization of the field. However, highly energetic modes will be less localized, as is well known that the region where the Hermite polynomials are non-negligible grows as $\sqrt{2n + 1}$. When we say that the field is localized for the harmonic potential, we mean that each of the relevant modes for a given physical scenario is indeed exponentially localized.

\bibliography{references.bib}

\end{document}